\documentclass[conference, a4paper]{IEEEtran}

\usepackage{graphicx}
\usepackage{amsmath, epsfig, amssymb, amstext, verbatim, amsopn, cite, subfigure, multirow, multicol, lipsum}
\usepackage{balance}
\usepackage{url}
\usepackage{amsfonts}
\usepackage{epsfig}
\usepackage{algorithmic}
\usepackage{algorithm}
\usepackage{setspace}
\usepackage{stmaryrd}
\usepackage{psfrag}	
\usepackage{multirow}
\usepackage{float}
\usepackage[process=auto]{pstool}
\usepackage{etoolbox}
\usepackage{color}
\usepackage{booktabs}
\usepackage{tabularx}

\allowdisplaybreaks
\usepackage{tikz}
\usetikzlibrary{arrows, decorations.markings}

\allowdisplaybreaks
\graphicspath{{images/}}

\newcommand{\vv}[1]{\boldsymbol{\mathrm{#1}}}
\newcommand{\mm}[1]{\boldsymbol{\mathrm{#1}}}
\newcommand{\expect}[1]{{\mathbb{E}}\!\left\{ #1 \right\}}
\newcommand{\herm}{{\sf{H}}}
\newcommand{\transp}{{\sf{T}}}
\renewcommand{\baselinestretch}{1.036}

\begin{document}
\IEEEoverridecommandlockouts

\title{Low-Complexity Hybrid Linear/Tomlinson-Harashima Precoding for Downlink Large-Scale MU-MIMO Systems\vspace*{-6mm}}
\author{\IEEEauthorblockN{ Shahram Zarei, Wolfgang Gerstacker, and Robert Schober}
\IEEEauthorblockA{Friedrich-Alexander-Universit\"at Erlangen-N\"urnberg, Erlangen, Germany \\
Email: \{shahram.zarei, wolfgang.gerstacker, robert.schober\}@fau.de\\
\thanks{This paper has been submitted for presentation at IEEE Global Communications Conference (Globecom) 2016.}
}}
    
\maketitle
%******************************************************************************
%
%******************************************************************************

\begin{abstract}
In this paper, we propose a novel low-complexity hybrid linear/Tomlinson-Harashima precoder (H-L-THP) for downlink large-scale multiuser multiple-input multiple-output (MU-MIMO) systems. The proposed precoder comprises an inner linear precoder which utilizes only the second order statistics of the channel state information (CSI) and outer THPs which use the instantaneous overall CSI of the cascade of the actual channel and the inner precoder. The user terminals are divided into groups, where for each group a THP successively mitigates the intra-group interference, whereas the inter-group interference is canceled by the inner linear precoder. Our simulation results show that the bit error rate (BER) of the proposed H-L-THP precoder is close to that of the conventional THP precoder, and is substantially lower than the BER of the commonly used regularized zero-forcing (RZF) precoder. Moreover, we derive exact expressions for the computational complexity of the proposed H-L-THP precoder in terms of the required numbers of floating-point operations. These results reveal that the proposed H-L-THP precoder has a much lower computational complexity than the conventional THP and RZF precoders, and is thus an excellent candidate for practical implementations.
\end{abstract}

%\IEEEpeerreviewmaketitle
%******************************************************************************
%
%******************************************************************************
\vspace*{2mm}
\section{Introduction}\label{Sec_Intro}
\IEEEPARstart{I}{n} recent years, multiuser multiple-input multiple-output (MU-MIMO) techniques have become a mature technology \cite{Spencer_ComMag2004, Gesbert07}. Today, MU-MIMO is a key element of many modern wireless communication standards such as Long Term Evolution Advanced (LTE Advanced). Using MU-MIMO both high power efficiency and high spectral efficiency can be achieved.

In this paper, we consider the downlink of a single-cell MU-MIMO system which embodies a vector Gaussian broadcast channel (GBC). It is well known that the capacity of the vector GBC can be achieved by nonlinear dirty paper coding (DPC) \cite{Caire2003, Viswanath2003}. However, DPC has a very high computational complexity and is hence infeasible for practical applications. 

Tomlinson-Harashima precoding is a nonlinear precoding scheme which achieves near-capacity performance. A Tomlinson-Harashima precoder (THP) was first proposed for temporal equalization, i.e., mitigation of intersymbol interference (ISI) in time-dispersive channels \cite{Tomlinson71, Harashima_TCOM72 }. The concept of Tomlinson-Harashima precoding for spatial equalization in MIMO systems was introduced in \cite{Fischer_THP_ITG2002, Fischer_ISIT2002}, and extended to frequency-selective MU-MIMO channels in \cite{Joham_THP_ITG_2004}. Although THP achieves near-capacity performance in MU-MIMO systems, its computational complexity is too high for many realistic practical scenarios, especially for base stations (BSs) with medium- or large-scale (massive) antenna arrays. 

On the other hand, linear precoders such as regularized zero-forcing (RZF) precoders have lower complexity, and provide good performance when the number of user terminals (UTs) is much smaller than the number of BS antennas \cite{Peel2005}. In particular, for a fixed number of UTs, if the number of BS antennas goes to infinity, RZF precoding becomes capacity achieving \cite{Rusek2013}. However, in scenarios, where the number of UTs is close to the number of BS antennas, linear precoders show poor performance, and are outperformed by their nonlinear counterparts. Nevertheless, their high computational complexity in medium- to large-scale MIMO systems prevents nonlinear precoders from being used in practice. This motivates us to propose a novel precoder which can serve a larger number of UTs than RZF precoders for a given number of BS antennas, and combines the benefits of the low-complexity of linear precoders and the high performance of nonlinear precoders.

A complexity-reduced \emph{linear} precoding scheme for large-scale MIMO systems was recently presented in \cite{Adhikary_TIT2013}. Here, the authors propose a per-group processing RZF (PGP-RZF) precoder which has a lower computational complexity than the conventional RZF precoder and provides a good compromise between performance and complexity when the number of UTs is much smaller than the number of BS antennas. In this scheme, UTs are assigned to groups such that the UTs in the same group have identical statistical channel state information (CSI). The PGP-RZF precoder comprises two components. With the first component, i.e., the inner linear precoder, which is solely based on statistical CSI, the UT groups are separated in the spatial domain. The second component, which depends on the instantaneous CSI, is composed of linear RZF precoders and mitigates the multiuser interference (MUI) in each group. Because of its linear structure, a major drawback of the PGP-RZF precoder is its low performance in scenarios, where the number of UTs is not much smaller than the number of BS antennas.

In this paper, we propose a hybrid linear/THP (H-L-THP) which performs precoding in two stages. In the first stage, a similar technique as in the first stage of the PGP-RZF scheme is used, where an inner linear precoder, which is solely based on statistical CSI, tries to block-diagonalize the channel matrix, i.e.,  minimize the inter-group interference. In the second stage, for each group, a THP successively eliminates the intra-group interference, i.e., the MUI between the UTs in the corresponding group. Since statistical CSI is almost static and changes very slowly over time, the inner precoder needs to be updated very infrequently, and only the outer per-group THPs have to be updated in every channel coherence interval. 
 
In contrast to \cite{Adhikary_TIT2013}, where linear RZF precoders are used to mitigate the MUI within the groups, in the proposed precoder, more advanced THPs are employed for intra-group MUI cancellation leading to a substantially higher performance, especially when the number of UTs is comparable to the number of BS antennas. In fact, our simulation results show that the performance achieved by the proposed H-L-THP is quite close to that of the conventional THP. Furthermore, we provide a computational complexity analysis for the proposed H-L-THP in terms of the required number of floating-point operations (FLOPs). Although in H-L-THP, in contrast to the PGP-RZF precoder, nonlinear per-group precoders are used, the computational complexity of the H-L-THP is only slightly higher than that of the PGP-RZF precoder. This is due to the fact that both the PGP-RZF precoder and the H-L-THP block-diagnoalize the channel matrix in their first stage, and this block-diagonalization is the computationally most expensive signal processing operation for both precoders. Moreover, in the proposed H-L-THP, the per-group THPs operate on square effective channel matrices having a much smaller size than the actual channel matrix, which leads to a radically reduced complexity of the H-L-THP compared to the conventional THP.

\emph{Notation:} Boldface lower and upper case letters represent column vectors and matrices, respectively. $\mm{I}_K$ denotes the $K \times K$ identity matrix and ${\left[\mm{A}\right]}_{k,:}$, ${\left[\mm{A}\right]}_{:,l}$, and ${\left[\mm{A}\right]}_{k, l}$ stand for the $k$th row, the $l$th column, and the element in the $k$th row and the $l$th column of matrix $\mm{A}$, respectively. $(\cdot)^*$ denotes the complex conjugate, and  $\mathrm{tr}(\cdot)$, $(\cdot)^{\transp}$, and $(\cdot)^{\herm}$ represent the trace, transpose, and Hermitian transpose of a matrix, respectively. $\expect{\cdot}$ stands for the expectation operator and $\mathcal{C} \mathcal{N} \left(\vv{u}, \mm{\Phi} \right)$ denotes a circular symmetric complex Gaussian distribution with mean vector $\vv{u}$ and covariance matrix $\mm{\Phi}$. Moreover, $\mathrm{diag}\left( a_1,\ldots,a_K \right)$ and $\mathrm{diag}\left( \mm{A}_1,\ldots,\mm{A}_K \right)$ denote a diagonal and a block-diagonal matrix with $a_1,\ldots,a_K$ and $\mm{A}_1,\ldots,\mm{A}_K$ on its main diagonal, respectively.
%******************************************************************************
%
%******************************************************************************
\vspace*{0mm}
\section{System Model and Benchmark Schemes}\label{Sec_SystemModel}
In this section, the considered system model is introduced and benchmark precoding schemes are presented.
\vspace*{1mm}
\subsection{System Model}\label{SubSec_SystemModel}
\vspace*{1mm}
We consider the downlink of a single-cell MU-MIMO system, where a BS with $N$ antennas simultaneously transmits data to $K$ single-antenna UTs. The load factor, i.e., the ratio of the number of UTs to the number of BS antennas is denoted by $\beta=K/N$. The data symbols intended for transmission to the UTs are stacked into vector $\vv{d} = [d_{1}, \ldots ,d_{K}]^\transp$, where $d_{k} \in \mathcal{A} = \left\lbrace a_\mathrm{I} + j a_\mathrm{Q} \ | \ a_\mathrm{I}, a_\mathrm{Q} \in \left\lbrace \pm 1, \pm 3, \ldots, \pm \left( \sqrt{M} -1 \right) \right\rbrace\right\rbrace$ is the $M$-QAM modulated data symbol of the $k$th UT, and $M$ is the modulation order and a square number. The vector of the stacked received signals of the UTs is given by
\vspace*{1mm}
\begin{align}
\vv{r}_\mathrm{x} = \mm{H}^\herm \vv{s}_\mathrm{x} + \vv{z},
\label{Eqn_DL_Sig_RX}
\end{align}
where $\mathrm{x} \in \left\lbrace \text{RZF}, \text{PGP-RZF}, \text{THP}, \text{H-L-THP} \right\rbrace$ denotes the type of the precoding scheme employed and $\mm{H} = \left[ \vv{h}_{1}, \ldots, \vv{h}_{K} \right] \in \mathbb{C}^{N \times K}$ is the channel matrix with $\mm{h}_{k}$ being the channel vector of the $k$th UT, $ k \in \left\lbrace1,\ldots,K\right\rbrace$. In this work, we assume a block flat fading channel and perfect CSI knowledge at the BS. We further assume a correlated channel model, i.e., $\vv{h}_{k} = \tilde{\mm{R}}_{k} \vv{\nu}_{k}$, where $\vv{\nu}_{k} \sim \mathcal{C} \mathcal{N} \left(\mm{0}, \mm{I}_N \right) $, and $\mm{R}_{k} = \mathbb{E}\lbrace \vv{h}_{k} \vv{h}^\herm_{k} \rbrace = \tilde{\mm{R}}_{k} \tilde{\mm{R}}^\herm_{k}$ represents the channel correlation matrix of the $k$th UT. Here, we adopt a uniform linear array (ULA) with a one-ring scattering model for the channel correlation, which was also employed in \cite{Wagner2012}. Accordingly, we have
\begin{align}
\left[\tilde{\mm{R}}_{k}\right]_{m, n}=  \frac{1}{\theta_{k, \mathrm{max}} - \theta_{k, \mathrm{min}}} \int_{\theta_{k, \mathrm{min}}}^{\theta_{k, \mathrm{max}}} e^{j 2 \pi \omega \left(m-n\right) } d \theta,
\end{align}
where we assume uniformly distributed angles of arrival (AoA). Here, $\theta_{k, \mathrm{min}}$ and $\theta_{k, \mathrm{max}}$ denote the minimum and the maximum angles of the physical paths corresponding to the $k$th UT, respectively. Moreover, $\omega$ is the normalized antenna spacing in wavelengths. In (\ref{Eqn_DL_Sig_RX}), $\vv{z}=\left[z_1,\ldots,z_K \right]^\transp \sim \mathcal{C} \mathcal{N} \left( \vv{0}, \sigma_z^2 \mm{I}_K \right) $ is the stacked vector of the additive white Gaussian noise (AWGN) samples of the $K$ UTs with $\sigma_z^2$ being the variance of the AWGN at the UTs. Here, we investigate the performance of the considered schemes with respect to the energy per information bit $E_\mathrm{b}$ divided by the one-sided noise power spectral density $N_0$, which is given by $E_\mathrm{b}/N_0 \triangleq P_\mathrm{TX}/ \left( K\sigma^2_z \log_2 \left( M\right) \right) $, where $P_\mathrm{TX}=\mathbb{E}\left\lbrace \vv{s}_\mathrm{x}^\herm \vv{s}_\mathrm{x} \right\rbrace$ is the average total transmit power.
%******************************************************************************
%
%******************************************************************************
\vspace*{0mm}
\subsection{Benchmark Precoding Schemes}\label{SubSec_Benchmarks}
The RZF precoder is among the most commonly used linear precoders for downlink MU-MIMO systems. Thus, in this paper, we adopt the RZF precoder as a benchmark scheme for the proposed H-L-THP. The transmit data vector generated by linear precoders is given by $\vv{s}_\mathrm{x}=\mm{V}_\mathrm{x} \vv{d}$, where $\mathrm{x} \in \left\lbrace \text{RZF}, \text{PGP-RZF} \right\rbrace$, and $\mm{V}_\mathrm{x} \in \mathbb{C}^{N \times K}$ is the precoding matrix which for the RZF precoder is given by \cite{Wagner2012}
\vspace*{1mm}
\begin{align}
%\mm{V}_\mathrm{BF} &= \zeta_\mathrm{BF} \mm{H} \label{Eqn_BF} \\
\mm{V}_\mathrm{RZF} &= \zeta_\mathrm{RZF} \mm{H} \left( {\mm{H}}^\herm {\mm{H}} + \frac{K \sigma^2_z }{P_\mathrm{TX}} \mm{I}_K \right)^{-1},
\label{Eqn_RZF}
\end{align}
where $\zeta_\mathrm{RZF}$ is a normalization factor which ensures that the constraint $\mathrm{tr}\left(\mm{V}_\mathrm{RZF} \mm{V}_{\mathrm{RZF}}^\herm \right)=K$ is met. Applying this constraint leads to an average total transmit power of $P_\mathrm{TX}=K P_s$, where $P_s$ is the variance of one QAM symbol.

The second benchmark scheme which we consider is the PGP-RZF precoder proposed in \cite{Adhikary_TIT2013}. Here, the authors consider $G$ groups of UTs, where in each group, there are $\bar{K}=K/G$ UTs having identical channel correlation matrices $\mm{R}_g$, where $g$ is the group index. The assumption that all UTs in one group have identical CSI statistics was made for the sake of simplicity in \cite{Adhikary_TIT2013}, and PGP-RZF was extended to the more realistic case, where the UTs in each group have similar but not necessarily identical statistical CSI in \cite{Nam_JSAC2014}. Accordingly, we have $\mm{H}= \left[ \tilde{\mm{H}}_1, \ldots, \tilde{\mm{H}}_G \right]$, where $\tilde{\mm{H}}_g = \left[ \vv{h}_{(g-1)\bar{K}+1}, \ldots, \vv{h}_{(g-1)\bar{K}+\bar{K}} \right] \in \mathbb{C}^{N\times\bar{K}}$ denotes the channel matrix of the UTs in group $g\in \left\lbrace1,\ldots,G\right\rbrace$. The PGP-RZF precoding matrix is given by $\mm{V}_\mathrm{PGP-RZF} = \mm{W} \mm{P}$, where $\mm{W}=\left[\mm{W}_1,\ldots,\mm{W}_G\right]$ and $\mm{P}=\mathrm{diag} \left( \mm{P}_1,\ldots,\mm{P}_G \right)$ with $\mm{W}_g \in \mathbb{C}^{N \times \bar{K}}$ and $\mm{P}_g \in \mathbb{C}^{\bar{K} \times \bar{K}}$ representing the inner and outer precoders for the $g$th group, respectively. The inner precoder $\mm{W}$ is a function of the CSI statistics only and it is designed for the minimization of the inter-group interference which is equivalent to the minimization of the off-diagonal elements of the cascade \cite{Adhikary_TIT2013}
\begin{align}
\mm{H}^\herm \mm{W} = \begin{bmatrix}
\tilde{\mm{H}}_1^\herm \mm{W}_1 & \tilde{\mm{H}}_1^\herm \mm{W}_2 & \cdots  \tilde{\mm{H}}_1^\herm \mm{W}_G \\
\tilde{\mm{H}}_2^\herm \mm{W}_1 & \tilde{\mm{H}}_2^\herm \mm{W}_2 & \cdots  \tilde{\mm{H}}_2^\herm \mm{W}_G \\
\vdots & \vdots &\vdots \\
\tilde{\mm{H}}_G^\herm \mm{W}_1 & \tilde{\mm{H}}_G^\herm \mm{W}_2 & \cdots  \tilde{\mm{H}}_G^\herm \mm{W}_G \\
\end{bmatrix}.
\label{Eqn_H_W}
\end{align}
Hence, the data symbols of the UTs in the $g$th group are ideally transmitted in the null space of the channel matrix of the UTs of all other groups, i.e., the groups with indices $\left\lbrace1,\ldots,g-1,g+1,\ldots,G\right\rbrace$. Accordingly, the following matrix is defined \cite{Spencer_TSP2004, Adhikary_TIT2013}
\begin{align}
\mm{\Psi}_g = \left[ \check{\mm{U}}_1,\ldots, \check{\mm{U}}_{g-1}, \check{\mm{U}}_{g+1}, \ldots, \check{\mm{U}}_{G} \right],
\end{align}
where $\check{\mm{U}}_{g} \in \mathbb{C}^{N\times L_g}$ contains the left eigenvectors corresponding to the $L_g$ dominant eigenvalues of $\mm{R}_g = \mm{U}_g \mm{\Sigma}_g \mm{U}^\herm_g$, where $\mm{U}_g$ and $\mm{\Sigma}_g$ are the matrix of the eigenvectors and the diagonal matrix of the eigenvalues of $\mm{R}_g$, respectively, obtained from singular value decomposition (SVD). Here, $L_g$ is a design parameter which should be optimized. We note that $\mm{\Psi}_g$ has rank $\sum_{g=1}^G L_g$. Hence, a unitary basis of the orthogonal complement of the space spanned by $\mm{\Psi}_g$, i.e.,  $\mathrm{Span} \left( \mm{\Psi}_g \right)$, is given by $\mm{E}^{(0)}_g$ which is a matrix containing the $N-\sum_{g^\prime=1, \ g^\prime \neq g}^G L_{g^\prime}$ rightmost columns of $\mm{\Phi}_g = \left[\mm{E}^{(1)}_g, \mm{E}^{(0)}_g\right]$, where $\mm{\Phi}_g$ is the matrix of the left eigenvectors of $\mm{\Psi}_g$. Finally, $\mm{W}_g$, i.e., the inner precoder of the $g$th group is calculated as the product of $\mm{E}^{(0)}_g$ and the $\bar{K}$ dominant eigenvectors of a matrix containing the projection of the channel vectors in group $g$ onto $\mm{E}^{(0)}_g$. Accordingly, we have $\mm{W}_g = \mm{E}^{(0)}_g \mm{A}^{(1)}_g$, where $\mm{A}^{(1)}_g$ is obtained from \cite{Adhikary_TIT2013}
\begin{align}
\mm{E}^{(0)}_g \mm{R}_g {\left( {\mm{E}^{(0)}_g}\right)}^\herm =\hspace*{-.5mm} \mm{E}^{(0)}_g \mm{U}_g \mm{\Sigma}_g \mm{U}^\herm_g {\left( {\mm{E}^{(0)}_g}\right)}^\herm \hspace*{-1mm}=\hspace*{-1mm} \mm{A}^{(1)}_g \mm{\Upsilon}_g {\left(  {\mm{A}^{(1)}_g} \right) }^\herm,
\end{align}
where $\mm{A}^{(1)}_g$ and $\mm{\Upsilon}_g$ are the matrix containing the left eigenvectors and the diagonal matrix of the eigenvalues of $\mm{E}^{(0)}_g \mm{R}_g {\left( {\mm{E}^{(0)}_g}\right)}^\herm$, respectively. The RZF precoding matrix for group $g$ is then given by \cite{Adhikary_TIT2013}
\begin{align}
\mm{P}_g = \zeta_g \check{\mm{H}}_g \left( \check{\mm{H}}_g^\herm \check{\mm{H}}_g + \frac{\bar{K} \sigma^2_z}{P_\mathrm{TX}} \mm{I}_{\bar{K}} \right)^{-1},
\end{align}
where $\check{\mm{H}}_g = \mm{W}_g^\herm \tilde{\mm{H}}_g$ is the effective channel matrix of the $g$th group, and $\zeta_g$ is a normalization factor which ensures that the transmit power constraint $\mathrm{tr} \left( \mm{W}_g \mm{P}_g \mm{P}_g^\herm \mm{W}_g^\herm \right)=\bar{K}$ is met \cite{Adhikary_TIT2013}. For the considered benchmark linear precoders, the $k$th element of the received vector $\vv{r}_\mathrm{x}$ is directly fed to the QAM demodulator of the $k$th UT.

\begin{figure}[tbp]
\begin{center}
\psfrag{d}[cc][cc][0.7]{$\vv{d}$}
\psfrag{xx}[cc][cc][0.7]{$\vv{x}$}
\psfrag{s_thp}[cc][cc][0.7]{$\vv{s}_\mathrm{THP}$}
\psfrag{B-I}[cc][cc][0.7]{$\mm{B}- \mm{I}_K$}
\psfrag{F}[cc][cc][0.7]{$\mm{F}$}
\psfrag{r1}[cc][cc][0.7]{${r_\mathrm{x}}_1$}
\psfrag{rK}[cc][cc][0.7]{${r_\mathrm{x}}_K$}
\psfrag{H}[cc][cc][0.7]{$\mm{H}^\herm$}
\psfrag{z1}[cc][cc][0.7]{$z_1$}
\psfrag{zK}[cc][cc][0.7]{$z_K$}
\psfrag{x1}[cc][cc][0.7]{$\xi_1$}
\psfrag{xK}[cc][cc][0.7]{$\xi_K$}
\psfrag{Mod M}[cc][cc][0.7]{$\mathrm{Mod}_M$}
\psfrag{dh1}[cc][cc][0.7]{$\hat{d}_1$}
\psfrag{dhK}[cc][cc][0.7]{$\hat{d}_K$}
\includegraphics[width=\linewidth, clip=true]{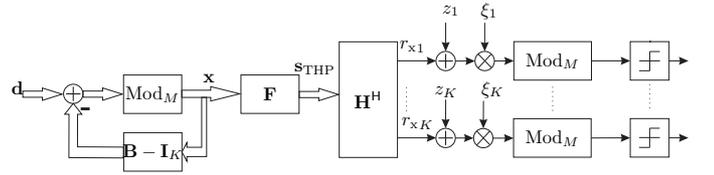}
\vspace*{0mm}
\caption{\linespread{0.9} \small System model for the conventional THP.}
\label{Fig_ConvTHP}
\end{center}
\end{figure}

The final benchmark scheme considered here is the conventional THP \cite{Fischer_THP_ITG2002, Fischer_ISIT2002}. The corresponding system model is shown in Fig. \ref{Fig_ConvTHP}. The THP transmit signal is given by $\vv{s}_\mathrm{THP}=\mm{F} \vv{x} $, where the unitary feedforward filter $\mm{F}$ is obtained from the QR-decomposition of the channel matrix $\mm{H}= \mm{F} \check{\mm{B}} $, and the elements of vector $\vv{x}$ are given by
\begin{align}
{\left[ \vv{x} \right]}_k \hspace*{-1mm} = \hspace*{-1mm} \mathrm{Mod}_M \left( {\left[ \vv{d} \right]}_k \hspace*{-2mm}-\hspace*{-1mm} \sum_{l=1}^{k-1} {\left[ \mm{B} \hspace*{-1mm}-\hspace*{-1mm} \mm{I}_{K} \right] }_{k, l} {\left[ \vv{x} \right]}_l \right), \forall k \in \left\lbrace 1, \ldots, K \right\rbrace,
\label{Eqn_tildex}
\end{align}
where the Modulo function $\mathrm{Mod}_M \left(x\right)$ is defined as
\begin{align}
\mathrm{Mod}_M \left(x\right) = &x - 2\sqrt{M}\bigg\lfloor \frac{1}{2} + \Re{ \left\lbrace \frac{x}{2\sqrt{M}} \right\rbrace } \bigg\rfloor \nonumber \\
&- 2j\sqrt{M} \bigg\lfloor \frac{1}{2} + \Im{ \left\lbrace \frac{x}{2\sqrt{M}} \right\rbrace } \bigg\rfloor
\label{Eqn_Modulo}
\end{align}
with $\Re{ \left\lbrace \cdot \right\rbrace}$ and $\Im{ \left\lbrace \cdot \right\rbrace}$ denoting the real and imaginary parts of a complex-valued variable. In (\ref{Eqn_tildex}), the feedback matrix $\mm{B}$ is given by $\mm{B}=\mm{\Xi}\check{\mm{B}}^\herm$, where $\mm{\Xi}=\mathrm{diag}\left(\xi_1,\ldots,\xi_K\right)$ with $\xi_k=1/[\check{\mm{B}}]_{k,k}, \forall k \in \left\lbrace 1,\ldots,K \right\rbrace$. At the $k$th UT, the received signal is first multiplied by $\xi_k$ and then passed through a $\mathrm{Mod}_M$-module before it is applied to the QAM demodulator. 
%******************************************************************************
%
%******************************************************************************
\vspace*{1mm}
\section{Hybrid Linear/THP}
\begin{figure}[tbp]
\begin{center}
\psfrag{d1}[cc][cc][0.7]{$\tilde{\vv{d}}_1$}
\psfrag{dG}[cc][cc][0.7]{$\tilde{\vv{d}}_G$}
\psfrag{y1}[cc][cc][0.7]{$\tilde{\vv{y}}_1$}
\psfrag{yG}[cc][cc][0.7]{$\tilde{\vv{y}}_G$}
\psfrag{s}[cc][cc][0.7]{$\vv{s}_\text{H-L-THP}$}
\psfrag{r1}[cc][cc][0.7]{${r_\mathrm{x}}_1$}
\psfrag{rK}[cc][cc][0.7]{${r_\mathrm{x}}_K$}
\psfrag{THP1}[cc][cc][0.7]{$\mathrm{THP}_1$}
\psfrag{THPG}[cc][cc][0.7]{$\mathrm{THP}_G$}
\psfrag{W}[cc][cc][0.7]{$\mm{W}$}
\psfrag{H}[cc][cc][0.7]{$\mm{H}^\herm$}
\psfrag{z1}[cc][cc][0.7]{$z_1$}
\psfrag{zK}[cc][cc][0.7]{$z_K$}
\psfrag{x1}[cc][cc][0.7]{$\xi_1$}
\psfrag{xK}[cc][cc][0.7]{$\xi_K$}
\psfrag{Mod M}[cc][cc][0.7]{$\mathrm{Mod}_M$}
\psfrag{dh1}[cc][cc][0.7]{$\hat{d}_1$}
\psfrag{dhK}[cc][cc][0.7]{$\hat{d}_K$}
\includegraphics[width=\linewidth, clip=true]{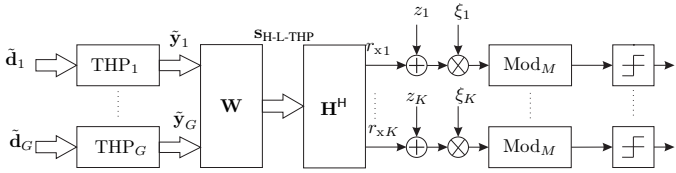}
\vspace*{1mm}
\caption{\linespread{0.9} \small System model for the H-L-THP.}
\label{Fig_SA_THP}
\end{center}
\end{figure}

In this section, the proposed H-L-THP is presented. The system model for H-L-THP is depicted in Fig. \ref{Fig_SA_THP}. Here, for the sake of simplicity, and in order to focus on the achievable performance gains and the main features of the proposed H-L-THP, we follow \cite{Adhikary_TIT2013} and assume that there are UTs that have identical channel correlation matrices. Furthermore, as in \cite{Adhikary_TIT2013}, we assume that the UTs with identical channel correlation matrices are assigned to the same group. \footnote{The more general case, where the UTs in a group have similar, but not necessarily identical CSI statistics, is left for future work. Considering the results for the PGP-RZF precoder in \cite{Nam_JSAC2014}, we do not expect that this generalization has a major impact on performance.} As can be seen from Fig. \ref{Fig_SA_THP}, H-L-THP is performed in two steps. In the first step, the linear precoder matrix $\mm{W}$ tries to minimize the inter-group interference, i.e., it transforms the matrix $\mm{H}^\herm$ into the semi-block-diagonal matrix $\mm{H}^\herm \mm{W}$ given in (\ref{Eqn_H_W}). In the second step, in each group $g$ with effective group channel matrix $\mm{W}_g^\herm \tilde{\mm{H}}_g \in \mathbb{C}^{\bar{K}\times \bar{K}}$, a THP module successively cancels the MUI interference. 

\begin{figure}[tbp]
\begin{center}
\psfrag{dg}[cc][cc][0.9]{$\tilde{\vv{d}}_g$}
\psfrag{Mod M}[cc][cc][0.9]{$\mathrm{Mod}_M$}
\psfrag{Bg-I}[cc][cc][0.9]{$\mm{B}_g - \mm{I}_{\bar{K}}$}
\psfrag{Fg}[cc][cc][0.9]{$\mm{F}_g$}
\psfrag{xg}[cc][cc][0.9]{$\tilde{\vv{x}}_g$}
\psfrag{yg}[cc][cc][0.9]{$\tilde{\vv{y}}_g$}
\includegraphics[width=0.8 \linewidth, clip=true]{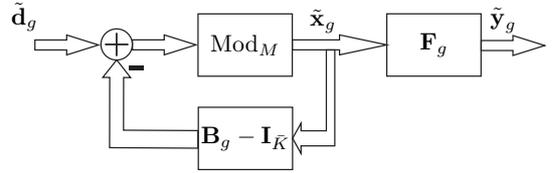}
\vspace*{0mm}
\caption{\linespread{0.9} \small System model for the $\mathrm{THP}_g$ module.}
\label{Fig_THP}
\end{center}
\end{figure}

For the block-diagonalization matrix $\mm{W}$, we employ the method of \cite{Adhikary_TIT2013} based on CSI statistics introduced in Section \ref{SubSec_Benchmarks}. Thereby, one key design parameter is the choice of suitable values for $L_g$. In particular, $L_g$ should be chosen in such a manner that the $G$ groups can be approximately separated in the signal space provided by the antenna array. Choosing values of $L_g$ that are too small results in a poor performance whereas choosing values for $L_g$ that are too large may lead to a situation, where only a small number of groups can be constructed, which will in turn result in larger group sizes and therefore a higher computational complexity. The optimization of $L_g$ is beyond the scope of this paper, and will be addressed in future work. The system model for the per-group THPs is shown in Fig. \ref{Fig_THP}. The input data vector of the THP module of the $g$th group is denoted by $\tilde{\vv{d}}_g \in \mathbb{C}^{\bar{K}\times 1}$, where $\vv{d} = \left[ \tilde{\vv{d}}_1^\transp,\ldots,\tilde{\vv{d}}_G^\transp \right]^\transp$. Similar to the conventional THP, the per-group THP module consists of a feedforward and a feedback part. The feedforward matrix $\mm{F}_g$ is obtained from the QR-decomposition of the effective group channel matrix $\mm{W}_g^\herm \tilde{\mm{H}}_g$ as follows
\begin{align}
\mm{F}_g \tilde{\mm{B}}_g = \mm{W}_g^\herm \tilde{\mm{H}}_g.
\end{align}
The feedback matrix $\mm{B}_g$ is then given by $\mm{B}_g = \mm{\Xi}_g \tilde{\mm{B}}_g^\herm$, where $\mm{\Xi}_g= \mathrm{diag} \left( \xi_{(g-1)\bar{K}+1},\ldots, \xi_{(g-1)\bar{K}+\bar{K}} \right)$ with $\xi_{(g-1)\bar{K}+k^\prime} = 1 / \left[ \tilde{\mm{B}}_g \right]_{k^\prime, k^\prime}, k^\prime \in \left\lbrace 1,\ldots,\bar{K} \right\rbrace$. The output data vector of the feedback part of the $g$th group, which is denoted by $\tilde{\vv{x}}_g$, is calculated according to
\begin{align}
{\left[ \tilde{\vv{x}}_g \right]}_{k^\prime} \hspace*{-1mm} = \hspace*{-1mm} \mathrm{Mod}_M \hspace*{-1mm} \left( \hspace*{-1mm} {\left[ \tilde{\vv{d}}_g \right]}_{k^\prime} \hspace*{-2mm}-\hspace*{-1mm} \sum_{l=1}^{k^\prime-1} {\left[ \mm{B}_g \hspace*{-1mm}-\hspace*{-1mm} \mm{I}_{\bar{K}} \right] }_{{k^\prime}, l} {\left[ \tilde{\vv{x}}_g \right]}_l \hspace*{-1mm} \right) \hspace*{-.5mm}
\label{Eqn_tildex_g}
\end{align}
for $k^\prime \in \left\lbrace 1, \ldots, \bar{K} \right\rbrace$. The output data vector of the THP module of the $g$th group is given by $\tilde{\vv{y}}_g = \mm{F}_g \tilde{\vv{x}}_g$. Finally, the transmit signal $\vv{s}_\mathrm{H-L-THP}$ is obtained as $\vv{s}_\mathrm{H-L-THP} = \sum_{g=1}^G \mm{W}_g \tilde{\vv{y}}_g$. At the $k^\prime$th UT in the $g$th group, the received signal is first multiplied by $\xi_{(g-1)\bar{K}+k^\prime}$ and then, after performing the same modulo operation as for the per-group THP modules, is fed to the QAM demodulator.

\vspace*{1mm}

\vspace*{0mm}

%******************************************************************************
%
%******************************************************************************
\vspace*{0mm}
\section{Computational Complexity Analysis}
In this section, we analyze the computational complexity of the proposed H-L-THP, and compare it to the computational complexity of the RZF precoder, the PGP-RZF precoder, and the conventional THP. Here, the computational complexity is expressed in terms of the required number of FLOPs corresponding to the number of complex-valued multiplications and additions. We assume that one complex-valued multiplication and one complex-valued addition require 6 and 2 FLOPs, respectively \cite{Arakava_TR2006}. Moreover, we are interested in the computational complexity required to generate $T$ precoded data vectors $\vv{s}_\mathrm{x}$, where $T$ is the length of the channel coherence interval, i.e., the number of the data symbols in a time interval, during which the channel does not change. Furthermore, we neglect the computational complexity of the SVDs required for the PGP-RZF precoder and the H-L-THP, since they have to be determined very infrequently as they are calculated based on statistical CSI. 

First, we determine the computational complexity of the RZF precoder. Generating the Gram matrix $\mm{H}^\herm \mm{H}$ requires $K(K+1)(4N-1)$ FLOPS, where the Hermitian property of the Gram matrix is exploited \cite{Arakava_TR2006, Hunger2007}. Adding the scaled identity matrix to $\mm{H}^\herm \mm{H}$ requires additional $K$ FLOPs, and taking the inverse of the resulting matrix requires $4K^3 + 8K^2 + 6K$ FLOPs \cite{Arakava_TR2006, Hunger2007}. For the multiplication of the inverse matrix with $\mm{H}$, $2NK(4K - 1)$ FLOPs are required, and $2NT(4K-1)$ FLOPs are required to generate $T$ precoded data vectors. This results in a total complexity of 
\begin{align}
C_\mathrm{RZF}= & 4K^3 + 2KN(4K - 1) + K(4N - 1)(K + 1) \nonumber \\
& + 8K^2+ 7K + 2NT(4K - 1).
\end{align}

Next, we derive the computational complexity of the PGP-RZF precoder. Calculation of $G$ effective per-group channel matrices $\check{\mm{H}}_g=\mm{W}_g^\herm \tilde{\mm{H}}_g$ requires $2G \bar{K}^2 \left( 4N-1 \right)$ FLOPs \cite{Arakava_TR2006, Hunger2007}. The derivation of the computational complexity of the remaining operations of the PGP-RZF precoder closely follows that of the RZF precoder, and we only present the total number of FLOPs, which is equal to 
\begin{align}
C_\mathrm{PGP-RZF}= & G\bar{K}(16 \bar{K}N + 16\bar{K}^2 + 7\bar{K} + 6 - 2N) \nonumber \\
& + 2NT(4K - 1),
\end{align}
where $\bar{K} = K/G$ is the number of UTs per group.

%\begin{table}[h!]
\begin{table*}[t]
  \centering
  \caption{Computational complexity of the considered precoding schemes.}
  \begin{tabular}{ p{8cm} p{8cm} }
    \toprule
    Precoding scheme & FLOPs\\
    \midrule
    RZF & $4K^3 + 2KN(4K - 1) + K(4N - 1)(K + 1) + 2NT(4K - 1)+ 8K^2+ 7K$\\
    \hline
    PGP-RZF & $ 2NT(4K-1) + 6K - 2KN + (16N + 16K/G + 7)K^2/G $\\
    \hline
    THP & $16K^3/3 + 2K(4KN-K+2) + 2T \big( 2K + 2K^2 + N \left( 4K - 1 \right) - 4 \big)$\\
    \hline
    
    H-L-THP & $2K(3G^2 - 3G^2N + 20K^2 - 6GK + 24GKN) / 3G^2  +  2T(2K^2 - 4G^2 + 2GK - GN + 4GKN) / G $\\
    \bottomrule
  \end{tabular}
\label{tab:table1}  
\end{table*}

Now, we derive the computational complexity of the conventional THP. The required number of FLOPs to perform QR-decomposition of channel matrix $\mm{H}$ is $8NK^2 - 8K^3/3$ \cite{Arakava_TR2006, Hunger2007}. Calculating matrix $\mm{B}-\mm{I}$ and generating $T$ data vectors at the output of the feedback part requires $8K^3 -2 K^2 + 2K$ and $4T(K^2+K-2)$ FLOPs, respectively \cite{Rodriguez_TCOM2014}. Finally, filtering the resulting vectors $T$ times with the feedforward part to generate $T$ precoded data vectors requires $2TN(4K-1)$ FLOPs. This results in a total number of FLOPs of
\begin{align}
C_\mathrm{THP} = & \frac{16K^3}{3} + 8K^2N - 2K^2 + 4K \nonumber \\
& + 2T \left( 2K + 2K^2 + N \left( 4K - 1 \right) - 4 \right)
\end{align}
FLOPs for the conventional THP.

Finally, we calculate the computational complexity of the H-L-THP. In the first step, $G$ matrix-matrix multiplications are performed to obtain the per-group effective channel matrices $\check{\mm{H}}_g = \mm{W}_g^\herm \tilde{\mm{H}}_g  \in \mathbb{C}^{\bar{K} \times \bar{K}}$, which require $2G \bar{K}^2 \left( 4N-1 \right)$ FLOPs \cite{Arakava_TR2006, Hunger2007}. The required number of FLOPs for performing $G$ QR-decompositions of the effective channel matrices $\check{\mm{H}}_g$ is $16 \ G \bar{K}^3 /3$. Calculating matrices $\mm{B}_g-\mm{I}_{\bar{K}}, g \in \left\lbrace 1,\ldots,G\right\rbrace$, and $TG$ computations of (\ref{Eqn_tildex_g}) require $2G\bar{K}\left(4 \bar{K}^2 - \bar{K} + 1\right)$ and $4TG( \bar{K}^2 + \bar{K} - 2)$ FLOPs, respectively \cite{Rodriguez_TCOM2014}. The required number of FLOPs for calculating $\tilde{\mm{W}}_g \mm{F}_g, g \in \left\lbrace 1,\ldots,G\right\rbrace$ is $2NG \bar{K}( 4 \bar{K} - 1 )$. Finally, $2TN(4K - 1)$ FLOPs are needed for generating $T$ precoded data vectors. This leads to the following total number of FLOPs for H-L-THP  
\begin{align}
C_\mathrm{H-L-THP}= & \frac{40G\bar{K}^3}{3} -4G\bar{K}^2  +  2G\bar{K}  - 2G\bar{K}N + 16G\bar{K}^2N  \nonumber\\
& + T(4G\bar{K}^2 + 4G\bar{K}+ 8KN- 8G- 2N).
\end{align}

The computational complexity of the considered precoders after substitution $\bar{K}=K/G$ in terms of the required number of FLOPs is summarized in Table \ref{tab:table1}.

%******************************************************************************
%
%******************************************************************************

\section{Numerical Results}\label{Sec_NumResults}
In order to evaluate the performance of the proposed H-L-THP, Monte-Carlo simulations have been performed. Here, we assume a single-cell system with a BS equipped with $N=32$ antennas transmitting data to $K$ single-antenna UTs. Moreover, for the antenna correlation, we adopt the same AoA model as in \cite{Adhikary_TIT2013}. Accordingly, we have $\theta_{g, \mathrm{min}} = -\pi + 2\pi(g-1)/G$ and $\theta_{g, \mathrm{max}} = -\pi + 2\pi(g-1)/G + 2\Delta$ for the $g$th group, where $\Delta$ is the angular spread of the UTs which is set to $10^\circ$. Furthermore, the normalized antenna spacing with respect to the wavelength is set to $\omega=0.5$. We adopt $M=16$ for the modulation order in all simulations. In this paper, we consider the average uncoded bit error rate (BER) as a performance metric, where the averaging is performed over a sufficient number of channel
realizations. We assume that the channel does not change within one data block, but changes from one block to the next independently.

\vspace*{1mm}
\begin{figure}[tbp]
\begin{center}
\psfrag{EbN0}[cc][cc][0.8]{$E_\mathrm{b}/N_0$ [dB]}
\includegraphics[width=\linewidth, clip=true]{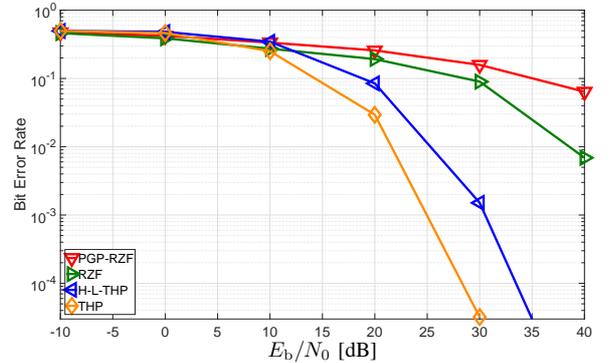}
\vspace*{1mm}
\caption{\linespread{0.9} \small Uncoded bit error rate vs. $E_\mathrm{b}/N_0$ for $N=32$, $K=16$, and $G=4$.}
\label{Fig_BER_K16}
\end{center}
\end{figure}

In Fig. \ref{Fig_BER_K16}, the average uncoded BER of the proposed H-L-THP is compared to that of the conventional THP, the RZF precoder, and the PGP-RZF precoder. The number of UTs and the number of groups is set to $K=16$ and $G=4$, respectively. As can be seen, for medium to high $E_\mathrm{b}/N_0$ values, THP achieves the lowest BER followed by the H-L-THP, the RZF precoder, and the PGP-RZF precoder. Moreover, from Fig. \ref{Fig_BER_K16}, it can be observed that the H-L-THP substantially outperforms the RZF and PGP-RZF precoders in the high $E_\mathrm{b}/N_0$ regime while having only a slightly higher computational complexity than the PGP-RZF precoder. This large performance gain of the H-L-THP compared to linear precoders comes from the more advance signal processing it performs at the BS whereas the poor performance of the linear precoders is due to the correlation of the channel vectors. The BER performance of the PGP-RZF precoder is worse than that of the RZF precoder which is the price paid for reducing the size of the effective channel matrix by exploiting statistical CSI in order to have a lower computational complexity. We note that this behavior is also reported in \cite{Adhikary_TIT2013}.

\begin{figure}[tbp]
\begin{center}
\psfrag{EbN0}[cc][cc][0.8]{$E_\mathrm{b}/N_0$ [dB]}
\includegraphics[width=\linewidth, clip=true]{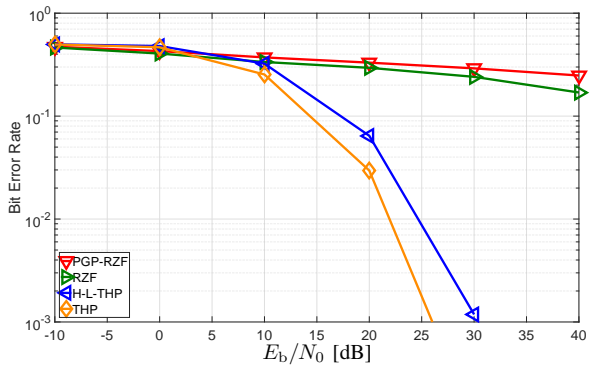}
\vspace*{1mm}
\caption{\linespread{0.9} \small Uncoded bit error rate vs. $E_\mathrm{b}/N_0$ for $N=32$, $K=32$, and $G=4$.}
\label{Fig_BER_K32}
\end{center}
\end{figure}

To further investigate the impact of the system load on performance, we present results for $K=N=32$ which corresponds to a load factor $\beta$ equal to one. For this simulation, the number of groups is also set to $G=4$. As can be seen from Fig. \ref{Fig_BER_K32}, in this case, the RZF and PGP-RZF precoders have a very poor BER performance, which is due to the fact that for $K=N$ there are not enough degrees of freedom to separate the UTs well in the spatial domain with linear precoding techniques. Both the THP and the H-L-THP achieve a considerably better performance than the linear precoders, since they employ a more sophisticated successive interference cancellation technique at the BS. From Fig. \ref{Fig_BER_K32}, it can also be seen that again the H-L-THP achieves a BER which is only slightly worse than that of the conventional THP while having a substantially lower computational complexity.

\begin{figure}[tbp]
\begin{center}
\psfrag{K}[cc][cc][0.8]{$K$}
\includegraphics[width=\linewidth, clip=true]{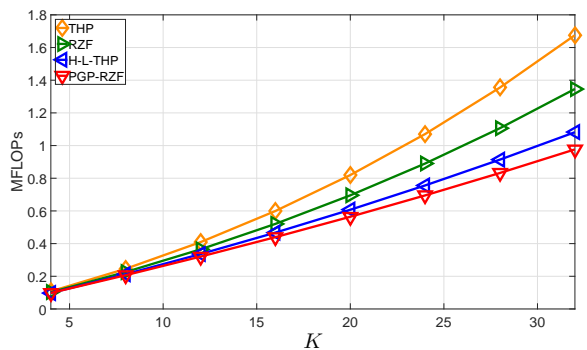}
\vspace*{1mm}
\caption{\linespread{0.9} \small Computational complexity vs. $K$ for $N=32$, $G=4$, and $T=100$.}
\label{Fig_Complexity}
\end{center}
\end{figure}

In Fig. \ref{Fig_Complexity}, we compare the computational complexities of the considered precoders in terms of the required numbers of Mega FLOPs (MFLOPs). As expected, THP has the highest computational complexity followed by the RZF precoder, the H-L-THP, and the PGP-RZF precoder. One important fact which can be observed in Fig. \ref{Fig_Complexity} is that the H-L-THP has only a slightly higher computational complexity than the PGP-RZF precoder. This is due to the fact that for both precoders the computation of the effective per-group channel matrices $\check{\mm{H}}_g=\mm{W}_g^\herm \tilde{\mm{H}}_g, g \in \left\lbrace 1, \ldots, G\right\rbrace$, is required and has a much higher computational complexity than all other operations needed for calculation of the PGP-RZF precoder and the H-L-THP. 

\vspace*{1mm}

\vspace*{1mm}

%******************************************************************************
%
%******************************************************************************
\section{Conclusion}\label{Sec_Conclusion}
We have presented a low-complexity H-L-THP scheme for MU-MIMO systems. The proposed H-L-THP achieves a similar BER performance as the conventional THP, and substantially outperforms the linear RZF and PGP-RZF precoders. We have also provided exact mathematical expressions for the computational complexity of the considered precoders in terms of the number of required FLOPs. Our complexity analysis has shown that, despite its excellent BER performance, the proposed H-L-THP has a considerably lower computational complexity than both the THP and the RZF precoder.
%******************************************************************************
%
%******************************************************************************
\vspace*{1mm}
\bibliographystyle{IEEEtran}
\bibliography{Massive_MIMO}
\end{document}